\documentclass[letterpaper]{article} % DO NOT CHANGE THIS
\usepackage{aaai25}  % DO NOT CHANGE THIS
\usepackage{times}  % DO NOT CHANGE THIS
\usepackage{helvet}  % DO NOT CHANGE THIS
\usepackage{courier}  % DO NOT CHANGE THIS
\usepackage[hyphens]{url}  % DO NOT CHANGE THIS
\usepackage{graphicx} % DO NOT CHANGE THIS
\usepackage{booktabs}
\usepackage{multirow}
\usepackage{multicol}
\usepackage{amsmath}
\usepackage{amssymb}
\usepackage{breqn}
\usepackage{threeparttable}
\usepackage{color}

\usepackage{natbib}  % DO NOT CHANGE THIS AND DO NOT ADD ANY OPTIONS TO IT
\usepackage{caption} % DO NOT CHANGE THIS AND DO NOT ADD ANY OPTIONS TO IT
\usepackage{algorithm}
\usepackage{algorithmic}

\usepackage{newfloat}
\usepackage{listings}
\DeclareCaptionStyle{ruled}{labelfont=normalfont,labelsep=colon,strut=off} % DO NOT CHANGE THIS
\floatstyle{ruled}
\newfloat{listing}{tb}{lst}{}
\floatname{listing}{Listing}
\nocopyright% -- Your paper will not be published if you use this command

\title{BLS-GAN: A Deep Layer Separation Framework for Eliminating Bone Overlap in Conventional Radiographs}
\author{
    Haolin Wang\textsuperscript{\rm 1}, 
    Yafei Ou\textsuperscript{\rm 2}\thanks{Corresponding Author: Yafei Ou (ou.y.ac@m.titech.ac.jp).},
    Prasoon Ambalathankandy\textsuperscript{\rm 3}, 
    Gen Ota\textsuperscript{\rm 4}, 
    Pengyu Dai\textsuperscript{\rm 2}, 
    Masayuki Ikebe\textsuperscript{\rm 4}, 
    Kenji Suzuki\textsuperscript{\rm 2}, 
    Tamotsu Kamishima\textsuperscript{\rm 5}
}
\affiliations{
    \textsuperscript{\rm 1}Graduate School of Health Sciences, Hokkaido University, Sapporo, Japan\\
    \textsuperscript{\rm 2}Institute of Innovative Research, Tokyo Institute of Technology, Yokohama, Japan\\
    \textsuperscript{\rm 3}Processer Research Team, RIKEN Center for Computational Science, Kobe, Japan\\
    \textsuperscript{\rm 4}Research Center For Integrated Quantum Electronics, Hokkaido University, Sapporo, Japan\\
    \textsuperscript{\rm 5}Faculty of Health Sciences, Hokkaido University, Sapporo, Japan\\
}
\usepackage{bibentry}

\begin{document}

\maketitle

\begin{abstract}
Conventional radiography is the widely used imaging technology in diagnosing, monitoring, and prognosticating musculoskeletal (MSK) diseases because of its easy availability, versatility, and cost-effectiveness.
In conventional radiographs, bone overlaps are prevalent, and can impede the accurate assessment of bone characteristics by radiologists or algorithms, posing significant challenges to conventional and computer-aided diagnoses.
This work initiated the study of a challenging scenario - bone layer separation in conventional radiographs, in which separate overlapped bone regions enable the independent assessment of the bone characteristics of each bone layer and lay the groundwork for MSK disease diagnosis and its automation.
This work proposed a Bone Layer Separation GAN (BLS-GAN) framework that can produce high-quality bone layer images with reasonable bone characteristics and texture. 
This framework introduced a reconstructor based on conventional radiography imaging principles, which achieved efficient reconstruction and mitigates the recurrent calculations and training instability issues caused by soft tissue in the overlapped regions. 
Additionally, pre-training with synthetic images was implemented to enhance the stability of both the training process and the results.
The generated images passed the visual Turing test, and improved performance in downstream tasks.
This work affirms the feasibility of extracting bone layer images from conventional radiographs, which holds promise for leveraging bone layer separation technology to facilitate more comprehensive analytical research in MSK diagnosis, monitoring, and prognosis. Code and dataset: \url{https://github.com/pokeblow/BLS-GAN.git}. 
\end{abstract}

\section{Introduction}
Conventional radiography (projection radiography) is a cost-effective and versatile diagnostic technology~\cite{pasveer1989knowledge, ou2021recent}, especially for the musculoskeletal (MSK) system ~\cite{grant2018musculoskeletal}, due to its ability to produce high-resolution and high-contrast bone images. However, a significant limitation of this modality is the bone overlaps \cite{newton2016handbook, low2017radiography}, which is prevalent in MSK imaging. This overlap in MSK radiographs introduces a textural mixture from both upper and lower tissues, complicating the accurate localization and analysis of bone lesions, eventually affecting the clinical diagnosis and management.

\begin{figure}[!t]
\centering
  \includegraphics[width=\linewidth]{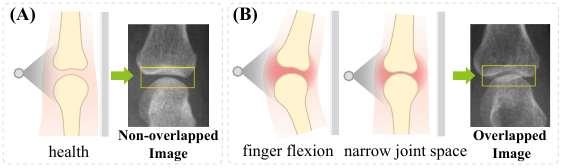}
  \caption{Explanation of bone overlap: Due to finger flexion and the excessively narrow joint space, the normal joint highlighted in the yellow box in (A) appears as the joint with bone overlap in (B). Bone overlap impedes clinical imaging diagnosis and its automatic analysis in MSK diseases.}
\label{fig: background}
\end{figure}

This work explores rheumatoid arthritis (RA), one of the prevalent MSK diseases. RA is a chronic autoimmune inflammatory disease characterized by joint swelling and tenderness. Physicians typically diagnose, prognose, and monitor RA by observing joint symptoms and imaging features, with joint space narrowing (JSN) and bone erosion in the fingers being crucial indicators for joint destruction~\citep{aletaha2018diagnosis, platten2017fully}. As the disease progresses, RA patients develop limited finger mobility, resulting from JSN, subluxation, and dislocation, which manifests in radiographic images as a transition from a clearly demarcated joint space with non-overlap to significant bone overlap, especially at the metacarpophalangeal (MCP) joints, as shown in Fig.~\ref{fig: background}~(A), (B). The texture mixture caused by the overlap poses a challenge to the imaging diagnosis of RA, especially the qualitative diagnosis and monitoring of JSN and bone erosion.
This also presents a new challenge for automating qualitative and quantitative analyses in RA, particularly in images with extensive overlap. For instance, bone overlap in automated JSN progression quantification methods not only reduces the accuracy and robustness of registration-based methods ~\cite{ou2022sub, wang2023deep} for some finger joint images, but also limits the scalability of these methods to other complex joints, such as the wrist, hip, and knee joints.

\begin{table}[!t]
    \centering
    \caption{Comparison of different scenario with proposed challenging scenario. G: Generator; D-like: Discriminator-like; R: Reconstruction. $\triangle$: represent a partial model involved.}
    \resizebox{\linewidth}{!}{
    \begin{tabular}{lcccc}
        \toprule
        \multirow{2}{*}{Scenario} & \multicolumn{3}{c}{Networks} & \multirow{2}{*}{Main Output} \\
        \cline{2-4}
        & G & D-like & R & \\
        \midrule
        Bone Suppression & $\checkmark$ & $\triangle$ & $\times$ & Soft tissue w/o bone \\
        Amodal Completion & $\checkmark$ & $\triangle$ & $\triangle$ & Occluded objects \\
        Image Removal & $\checkmark$ & $\triangle$ & $\checkmark$ & Image w/o objects \\
        BLS (Ours) & $\checkmark$ & $\checkmark$ & $\checkmark$ & Bone images w/o overlaps \\
        \bottomrule
    \end{tabular}
    }
    \label{tab: task}
\end{table}

In related works, amodal completion with inpainting has been widely employed to reconstruct occluded regions in natural images, achieving notable advances ~\cite{ao2023image, zhang2023image, Sargsyan_2023_ICCV, ko2023continuously, xu2024amodal}. 
Image removal can successfully remove impurities such as shadows and raindrops by reconstructing background textures ~\cite{elad2023image}. 
In chest radiography, rib suppression leveraging deep learning models have effectively removed the ribs ~\cite{suzuki2006image, han2022gan}, thereby enhancing the visibility of lung soft tissues and improving diagnostic efficiency for lesions. 
As illustrated in Table~\ref{tab: task}, there are substantial differences between our challenging scenario and others in terms of network structure design and output.
According to the imaging principle of conventional radiography ~\cite{bushberg2011essential}, the images show the superposition of the X-ray absorption rates by different tissues, leading to significant differences between conventional radiographs and natural images. Although parallels can be drawn in terms of scenario descriptions and application contexts with amodal completion, the lack of a robust reconstruction mechanism results in discrepancies between the generated textures and authentic bone textures. Moreover, the complex texture characteristics of bones set them apart from physical artifacts, such as shadows, rendering classical image removal methods less effective in this challenging scenario. In the context of bone (rib) suppression, existing methods are constrained by their limited ability to achieve distinct separation of the bone layer, and they fail to produce the desired outcomes.

To address these challenges, we initiate a challenging research scenario - joint bone layer separation, and propose a multi-supervised framework named \textbf{B}one \textbf{L}ayer \textbf{S}egmentation \textbf{G}enerative \textbf{A}dversarial  \textbf{N}etwork (\textbf{BLS-GAN}), which implements separated bone layer images extraction from a single finger conventional radiographs and eliminates bone overlap in each bone layer image. This generative-based method provides a reliable image basis for the independent evaluation of each bone layer feature and the study of automated analysis methods. Specifically, our contributions can be summarized as follows:
\begin{itemize}
\item 
\textbf{A Challenging Scenario for Amodal Completion}: The imaging principles of conventional radiography inherently result in bone overlap, which poses significant challenges for the clinical diagnosis and analysis of MSK system lesions, as well as for the development of automated qualitative and quantitative analysis methods. This issue is particularly problematic for diseases such as RA. The presence of overlaps can lead to substantial inaccuracies in the downstream JSN quantification task. Additionally, due to the requirement for strict adherence to the original texture of bones, classical amodal completion with inpainting are not feasible. This inspiration has spurred the exploration of a challenging research scenario. 
\item 
\textbf{Bone Layer Separation Framework}:
This work designed and implemented a novel framework for the above challenging scenario. Compared with other methods, our framework offers the following two innovations. (1) introduced a radiography imaging principles-based reconstructor that leverages conventional radiography principles and includes a correction parameter to rectify overlapped regions in the reconstruction. (2) integrated a segmentation-based multi-channel supervisor network to distinguish between overlapped and non-overlapped regions, enhancing the authenticity and natural appearance of bone textures in the generated images.

\item 
\textbf{Expert Assessments and Clinical Downstream Validation}: This work successfully passed the radiological technologist visual Turing test and can significantly enhance both the accuracy and stability in the clinical downstream task of JSN quantification.

\item 
\textbf{Dataset for this Challenging Scenario}: We provide a dataset specifically designed for this challenging scenario, utilized in this paper. The dataset includes joint images and the mask annotations for the upper and lower bones.

\end{itemize}

\begin{figure*}[!t]
\centering
  \includegraphics[width=\textwidth]{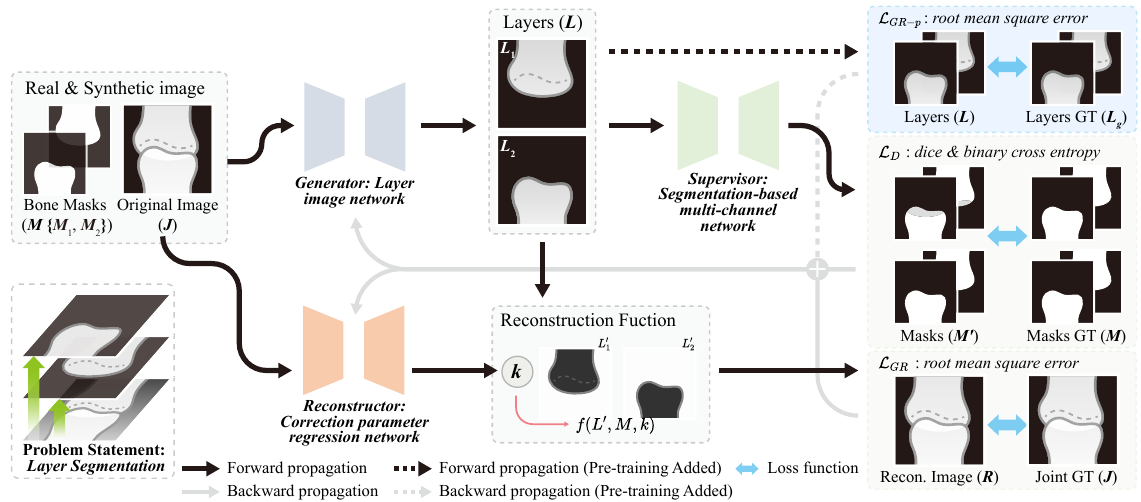}
  \caption{Explanation of bone layer separation: Layer-by-layer extraction of the upper and lower bones, followed by eliminating overlapped regions. The framework consists of three primary components: a generator, a supervisor, and a reconstructor. This process is performed as follows: (i) The generator produces bone layer images using the original joint image and corresponding bone masks as input. (ii) The layer images are discriminated by a segmentation-based multi-channel network and reconstructed through reconstructor, yielding a discrimination mask and a reconstructed image. (iii) Discrepancies between the masks and ground truth (GT), and between the reconstructed and original images, are used to create a hybrid loss function that guides the generator and reconstructor during back propagation. (iv) In the training pipeline, pre-training is preformed in synthetic images, the discrepancies of the real bone layer images is incorporated into the original loss function to facilitate the establishment of the initial model. Subsequently, the training is performed in real and synthetic images.}
\label{fig: network}
\end{figure*}

\section{Methodology}
Conventional hand radiographs of patients with RA often suffer from bone overlap in finger joints due to disease or positioning, leading to the mixture of texture information between the upper and lower bones. This poses several clinical and technical challenges. Thus, we explored a challenging research scenario: using single-layer conventional finger joint radiography as input,  generating independent layer images of the upper and lower bone without overlap as output, as illustrated in Fig.~\ref{fig: network}. We define this process as \textit{Bone Layer Separation}.

\subsection{Bone Layer Separation Framework}
This work proposed a GAN-based bone layer separation framework for finger joint radiographs to extract layer images and eliminate bone overlap in each layer image. As shown in Fig.~\ref{fig: network}, the framework consists of three basic sub-networks: the layer image generator, the segmentation-based multi-channel supervisor, the correction parameter regression reconstructor, and synthetic images pre-training. We define that bone layer separation of the original image, layer images \(1, ..., i, ..., n\) are generated. This study utilizes MCP joint images, therefore, $n$ is set to 2.

\paragraph{Layer Image Generator}
In the original image, we defined and partitioned the upper and lower bones of the joint as two independent layers. However, in the presence of bone overlap, each layer contains regions that intersect with other bone layers. The goal of the generator is to eliminate these overlapped regions within each bone layer, thereby achieving the separation defined in our study. The backbone network employed in this context can be any generation network.

The input is the original joint image \(J\) and its corresponding masks delineating the upper and lower bones of the joint, denoted as \(M = \{M_1, ..., M_n\}\). The output of the generator network is the layer images of the upper and lower bones in the joint with masks, defined as \(L = \{L_1, ..., L_n\}\). Assuming that the layer image generator is denoted as \(\mathcal{G}\), the generation process can be defined in Eq.~\ref{eq:generator}.
\begin{equation}
L = \mathcal{G}(J, M)\cdot M
\label{eq:generator}
\end{equation}

\paragraph{Segmentation-based Multi-channel Supervisor} Unlike traditional discriminators in GANs, which distinguish between real and fake images for the entire image, we integrated an segmentation multi-channel network to achieve pixel-level discrimination of layer images, named supervisor. Our supervisor outputs two sets of four-channel masks. One set pertains to the segmentation of overlapped and non-overlapped regions, and the other set assesses the authenticity of the generated images. Therefore, our segmentation-based supervisor effectively identifies overlapped regions in images, thereby enhancing generator supervision. The backbone network employed in this context can be any segmentation network.

The layer images from generator $L$ serve as the input to the supervisor. The output of the supervisor is defined as \(M' = \{M'_1, ..., M'_n\}\). Suppose the supervisor network is denoted as \(\mathcal{D}\). Therefore, the discrimination process can be defined as Eq.~\ref{eq:supervisor}, where $M'$ represents the discrimination mask from the supervisor. 
\begin{equation}
M' = \mathcal{D}(L) \cdot M
\label{eq:supervisor}
\end{equation}

\paragraph{Radiography Imaging Principles based Reconstructor}
According to the principles of conventional radiography, different tissues exhibit varying absorption rates. Tissues with higher density demonstrate greater absorption, while those with lower density exhibit weaker absorption, thereby giving rise to radiographic representations ~\cite{bushberg2011essential, huda2015radiographic}. In the presence of tissue overlap, X-ray absorption by the upper layer tissues influences the imaging of overlapped tissues, showing an exponential decay.

In contrast to amodal completion in natural images, adherence to the imaging principles of radiography is crucial. Therefore, we introduce a reconstructor to thoroughly supervise layer image generation. In this reconstructor, the absorption rate image of the bone is calculated based on the layer image, and reconstructed based on the reconstruction function defined below.
Our framework delineates the layer image of bones as a composite of bone texture and soft tissue texture. However, this amalgamation leads to the recurrent calculation of soft tissue texture within overlapped regions, thereby compromising reconstruction quality. To address this issue, we introduce a single correction parameter regression network using the VGG-18 network~\cite{simonyan2014very}, named correction parameter regression network, to derive a correction parameter to mitigate the impact of redundant soft tissue calculations within overlapped regions.

The algorithm flow proceeds as follows: Suppose \(\mathcal{R}\) denoted as the reconstructor, with the original joint image as input and a single parameter $k$ as output, which can be defined as \(k = \mathcal{R}(I)\). 
Subsequently, the image is reconstructed according to the reconstruction function \(f(L, M, k)\), as delineated in Eq.~\ref{eq:reconstructor}, where \( R \) denotes the reconstructed image and \(M_\cup = \bigcup_{i=1}^{n} M_i \). Specifically, the bone absorption rate image $L'$  can be defined as \(L' = 1 - L\). 
In the mask regions corresponding to the upper and lower bones of the generated layer image, divide by the correction parameter \( k \), followed by multiplying the layer superposition results in the mask regions by \( k \).
\begin{equation}
\begin{split}
R = f(L',M,k)= \left(1- k\prod_{i=1}^n(1-\left(1-\frac{L'_i}{k}\right) \cdot  M_i)\right) \cdot M_\cup
\end{split}
\label{eq:reconstructor}
\end{equation}

\paragraph{Framework Flowchart}
The image size processed by our framework is set to 256 $\times$ 256. We construct the loss function based on binary cross entropy (BCE) dice loss \(\mathcal{L}_{b}\) \cite{yeung2022unified} and root mean squared error (RMSE) loss \(\mathcal{L}_{r}\) \cite{chai2014root}.

We designate the GT as \({J}\) and \(M\). For the supervision of the generator and reconstructor, we employ the loss functions \(\mathcal{L}_b\) and \(\mathcal{L}_r\). Furthermore, we incorporate loss supervision additionally for the overlapped regions. Thus, the loss function of the networks can be defined as Eq. \ref{eq:loss}, where \(M_\cap = \bigcap_{i=1}^{n} M_i \), \(J_\cap = J \cdot M_\cap\), \(R_\cap = R \cdot M_\cap\),  \(\alpha_0 = \beta_0 = 0.5\).
\begin{equation}
\mathcal{L}_{GR} = \alpha_0 \times \mathcal{L}_{b}(M', M) + 
\beta_0 \times \mathcal{L}_{r}(R, J) + \mathcal{L}_{r}(R_\cap, J_\cap)
\label{eq:loss}
\end{equation}

In addition, we train the supervisor simultaneously and independently. Regarding the input for the supervisor, the real sample comprises the original image with the mask, denoted as $J_r = J \cdot M$ where $J$ represents the original joint image and $M$ denotes the masks. The GT of real samples \(M_r =\left \{\left \{M_1 - M_\cap, ..., M_n - M_\cap \right \}, \left \{M_1, ..., M_n \right \}\right \}\) is derived by eliminating the masked regions.
Conversely, the fake sample consists of the layer image generated by the generator, expressed \(J_f = \mathcal{G}(J, M)\). The GT for fake samples \(M_f =\left \{\left \{M_1, ..., M_n \right \}, \left \{\mathbf{0}_1, ..., \mathbf{0}_n  \right \}\right \}\), where \(\mathbf{0}\) represents an all-zero matrix. We performed \(\mathcal{L}_0\) loss for supervisor supervision. Thus, the loss function can be defined in Eq.~\ref{eq:disloss}, where $\alpha_1$ and $\beta_1$ are set to 0.5.
\begin{equation}
\mathcal{L}_D = \alpha_1 \times \mathcal{L}_b(\mathcal{D}(J_r), M_r) + \beta_1 \times \mathcal{L}_b(\mathcal{D}(J_f), M_f)
\label{eq:disloss}
\end{equation}

\paragraph{Synthetic Images Pre-training}
We constructed synthetic images with overlap based on images with non-overlap. Specifically, we utilize the non-overlap real image as the foundation and randomly shift the upper and lower articular bones to create an overlapped region. Reconstruction is subsequently performed based on the upper and lower bones using reconstruction function Eq.~\ref{eq:reconstructor} to generate the overlapped region. Regarding correction parameter $k$ in synthetic images, we utilized mathematical method for its determination. This process involves initially excluding the bone region from the image, subsequently solving the Laplace equation \cite{gong2020decompose}, and ultimately calculating the mean value within the overlapping region.

We process pre-training the utilizes synthetic images \(S\), specifically, since synthetic images are generated from non-overlap images, upper and lower bone GT \(L_g\) can be effectively obtained. Therefore, we introduce \(L_g\) into the loss function as defined in Eq. \ref{eq:pregen} and \ref{eq:predis}, to build initial capabilities of the framework. 

We performed pre-training of our framework with synthetic images and their corresponding masks to establish its foundational functionality. Specifically, since the synthetic images are synthesis from non-overlap images, we can effectively obtain the upper and lower bone GT (bone with non-overlap) \(L_g\) with non-overlap and its corresponding mask \(M\). Consequently, \(L_g\) is incorporated into Eq. \ref{eq:pregen} as the GT for loss function calculation. Additionally, \(L_g\) is introduced as the additional real sample of the supervisor in training, and loss function is presented in Eq. \ref{eq:predis}, where \(M_g = \{M, M\}\).
\begin{equation}
\mathcal{L}_{GR-p} = \mathcal{L}_{GR} + \mathcal{L}_r(L, L_g)
\label{eq:pregen}
\end{equation}
\begin{equation}
\mathcal{L}_{D-p} = \mathcal{L}_{D} + \mathcal{L}_b(\mathcal{D}(L_g), M_g)
\label{eq:predis}
\end{equation}

In the main training, due to the absence of GT for the upper and lower bone in real images, we continue to apply the original loss function Eq. \ref{eq:loss} and Eq. \ref{eq:disloss}, and framework trained employed both synthetic and real images.

\subsection{Implementation}
The networks were implemented on a workstation with three GPUs (NVIDIA GeForce GTX 2080 Ti). The supervisor, generator, and reconstructor networks were trained using the Adam optimizer with an initial learning rate of \(1e^{-5} \).
In our practice, we commence by performing pre-training on synthetic images and their corresponding GT images, extending this preparatory phase across 250 epochs with a consistently maintained batch size of 12. Subsequently, we refine the loss function and GT, maintaining the same batch size, for an additional 50 epochs with both synthetic and real images to meticulously optimize the performance of our framework. 
Considering the randomness of GAN networks, all networks are trained three times and retain the best parameters.

\section{Experiments}
To validate the robustness and reliability of the framework, we designed and conducted experiments to rigorously assess the fidelity of the generated layer images. 
Expert assessments and clinical validation were conducted to ascertain the performance and clinical relevance of this work. Comparison experiments and an ablation study were conducted to underscore the importance of our framework for this challenging scenario and the necessity of the network architecture. We concentrated primarily on the MCP joint due to its critical significance in disease diagnosis and its higher susceptibility to overlap in practice, which presents more significant challenges in the quantification of JSN progression and the development of automated algorithms.

We evaluated the reconstructed images with real overlap images. These are four metrics used in our experiments: the mean squared error (MSE), the structural similarity (SSIM), the peak signal-to-noise ratio (PSNR), and fréchet inception distance (FID).

\subsection{Dataset}
We prepared a clinical data set in accordance with the Declaration of Helsinki and obtained approval from the Ethics Committee of *** ***. 
% of the Faculty of Health Sciences, Hokkaido University.
The dataset utilized in this study comprises 168 posteroanterior (PA) radiographs of the hand sourced from 43 patients with RA. Of these patients, 88.5\% are female. The average age in the dataset is 65.6 years, with a variance of 12.87 and an age range of 31-91 years. These images originate from the *** ***, which employs its proprietary conventional radiography system and adheres to the Digital Imaging and Communications in Medicine (DICOM) standard for dataset management. Digital radiographs were acquired with the CALNEO smart C47 (Fujifilm, Tokyo, Japan) under the following conditions: tube voltage of 50 kV, tube current of 100 mA, exposure time of 0.02 milliseconds, source-to-image distance of 100 cm, resolution of 0.15 mm/pixel, image size of 1670$\times$2010 pixels, and a bit depth of 16 bits. Given that MCP joints are more susceptible to bone overlap than other joints, our dataset exclusively utilized MCP joint images.
These images were manually screened to remove those exhibiting severe bone erosion, resulting in a total of 1,594 joint images, which included 672 overlapping images, with an average overlap size of 686.05 ± 1983.12 pixels.
The dataset contains 430 MCP joints, which are divided into training and test sets by joint at a ratio of 3:1.
An experienced radiological laboratory assistant annotated these images to label the upper and lower bones of the joint into two channels, which were then further reviewed by a radiologist.

\begin{figure}[!t]
  \includegraphics[width=\linewidth]{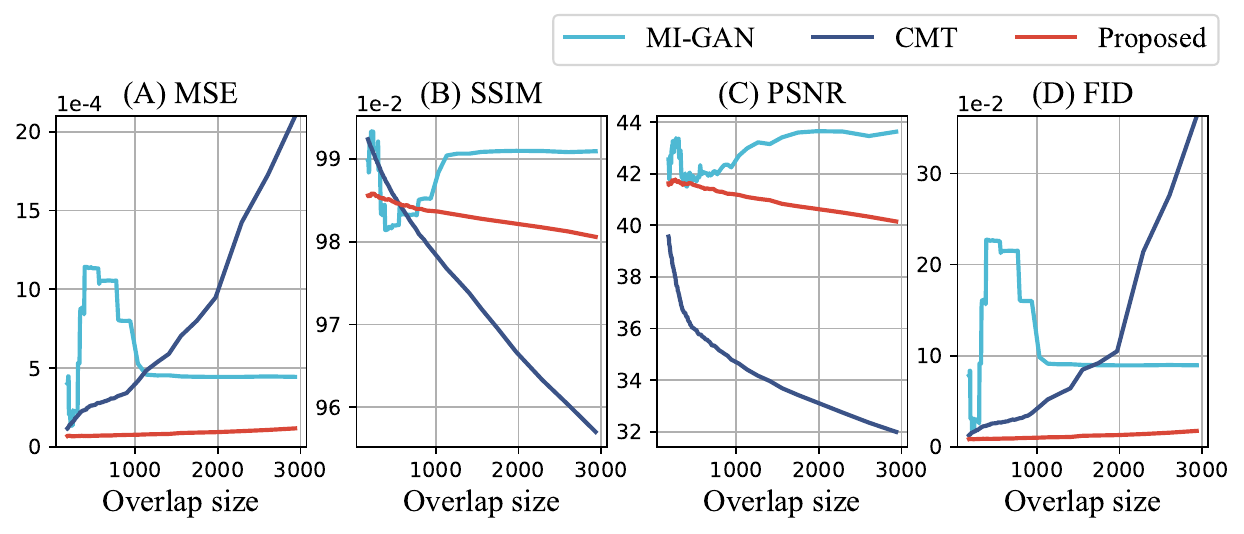}
  \centering
  \caption{Comparison of proposed framework with other methods in different metrics across overlap sizes.}
\label{fig: compare_plot}
\end{figure}

\begin{figure}[!t]
  \includegraphics[width=\linewidth]{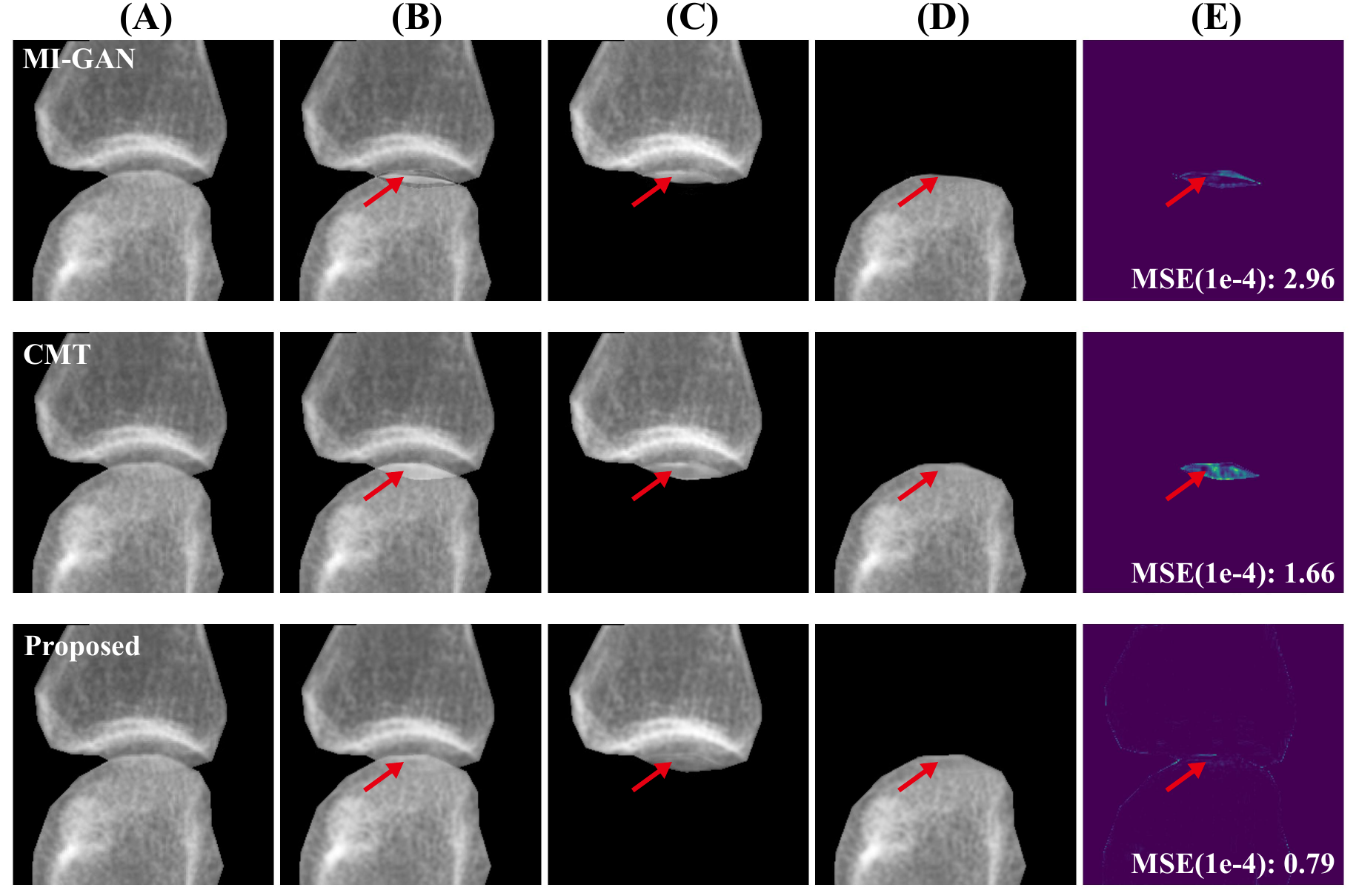}
  \centering
  \caption{Comparison of proposed framework with other methods. (A) Real Joint image; (B) Reconstructed Joint Image; (C) Upper Bone Layer; (D) Upper Bone Layer; (E) MSE Spectrum (A v.s. B).}
\label{fig: compare_image}
\end{figure}

\begin{table}[!t]
\caption{Evaluation result of proposed framework and comparison 
 with other methods in different metrics. Expressed as mean $\pm$ standard deviation.}
\centering
\resizebox{\linewidth}{!}{
    \begin{tabular}{lcccc}
        \toprule
        Method & MSE (\(10^{-4}\)) & SSIM (\(10^{-2}\)) & PSNR & FID (\(10^{-2}\)) \\
        \midrule
        MIN-GAN & 5.63$\pm$23.72 & \textbf{98.82$\pm$3.05} & \textbf{42.63$\pm$6.70} & 10.70$\pm$50.78 \\
        CMT & 9.26$\pm$32.27 & 97.83$\pm$3.28 & 35.94$\pm$4.83 & 15.83$\pm$70.78 \\
        Proposed & \textbf{0.88}$\pm$\textbf{0.70} & 98.37$\pm$0.51 & 41.09$\pm$1.72 & \textbf{1.20}$\pm$\textbf{1.33} \\
        \bottomrule
    \end{tabular}
}
\label{tab: methodcompare}
\end{table}

\subsection{Generate Image Evaluation}
We conducted experiment to evaluate the performance of our framework and compare with other amodal completion methods with inpainting. The evaluation was performed on real images with overlaps. For the amodal completion network with inpainting, we implemented the model in \cite{Sargsyan_2023_ICCV} and \cite{ko2023continuously}. Utilizing the pre-training parameters, we subsequently trained on our dataset. The network independently predicts the upper and lower bones and reconstructed using the reconstruction function of synthetic images.

As shown in Table~\ref{tab: methodcompare}, the generated layer images after reconstruction using our framework demonstrate exceptional performance across all four evaluation metrics. Our framework exhibits high accuracy and reliability compared to other methods, as shown by a significantly lower average MSE and FID. Furthermore, In Fig.~\ref{fig: compare_plot}, as the size of the overlapped regions increases, the evaluation index declines. This is because in cases with large overlap sizes, the joint generally suffers from severe bone erosion and extreme narrowing of the joint space, which leads to notable changes in the bone texture. Consequently, compared to cases with other overlap sizes, the task complexity for the framework increases significantly. Nonetheless, our framework continues to significantly outperform other methods, particularly demonstrating greater robustness at larger overlap sizes.

Additionally, as shown in Fig. \ref{fig: compare_image}, the bone layer separation framework achieves the extraction of bone images in the layer from a single image, effectively eliminating overlapped regions and preserving the complete bone texture. The reconstructed images closely resemble the original images, demonstrating excellent generation quality, which is further corroborated by the loss spectrum diagram. 
In overlap situations, our framework effectively eliminates small overlaps, though the perceptual impact may be subtle. As overlap increases, our refined elimination method becomes more effective. Overlapped regions retain the bone textural characteristics, ensuring smooth continuity between overlapped and non-overlapped areas. However, in cases of large overlaps, such as in joints with severe bone erosion, the generation of overlapped regions can be unstable, with noticeable sharp edges. Despite this, we successfully extracted layer images and eliminated bone overlap.
Compared to other methods, our framework exhibits significant superiority in both layer image generation and reconstruction, particularly in large overlap, thereby showcasing outstanding accuracy and robustness. 

In the amodal completion method with inpainting, the absence of a reconstruction process and non-adherence to conventional radiography principles result in an inability to accurately generate the texture of overlapped regions, leading to reduced accuracy and robustness. Conversely, our framework adheres strictly to imaging principles and incorporates supervision within the reconstruction , thereby enabling precise generation of textures in overlapped regions. Particularly in large overlap, where the generation based on amodal completion, lacking real image supervision, exhibits limitation of generation. In contrast, our framework integrates the synthetic images pre-training, the segmentation based supervisor and reconstruction structure, facilitating unsupervised network training on real images and extending the performance to accommodate a broader range of overlap.

\begin{figure}[!t]
  \includegraphics[width=\linewidth]{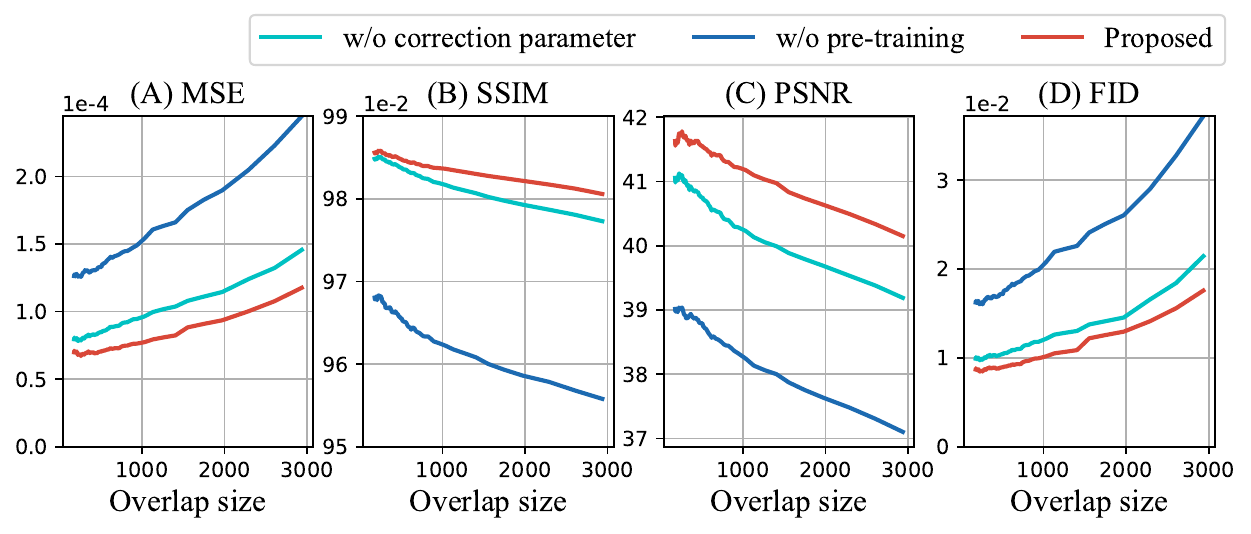}
  \centering
  \caption{Ablation study results of our framework in different metrics across overlap sizes.}
\label{fig: ablation_plot}
\end{figure}

\begin{table}[!t]
\caption{Comparison results in ablation study.}
\centering
\resizebox{\linewidth}{!}{
    \begin{tabular}{cccccc}
        \toprule
        P & C & MSE (\(10^{-4}\)) & SSIM (\(10^{-2}\)) & PSNR & FID (\(10^{-2}\)) \\
        \midrule
        & \checkmark & 1.72$\pm$1.62 & 96.33$\pm$1.10 & 38.30$\pm$1.86 & 2.46$\pm$3.26 \\
        \checkmark & & 1.05$\pm$0.88 & 98.20$\pm$0.70 & 40.34$\pm$1.76 & 1.44$\pm$1.86 \\
        \checkmark & \checkmark & \textbf{0.88}$\pm$\textbf{0.70} & \textbf{98.37}$\pm$\textbf{0.51} & \textbf{41.09}$\pm$\textbf{1
        .72} & \textbf{1.20}$\pm$\textbf{1.33} \\
        \bottomrule
    \end{tabular}
    }
\begin{tablenotes}
\item P: Pre-training, C: Correction Parameter.
\end{tablenotes}
\label{tab: ablation}
\end{table}

\begin{figure}[!t]
  \includegraphics[width=\linewidth]{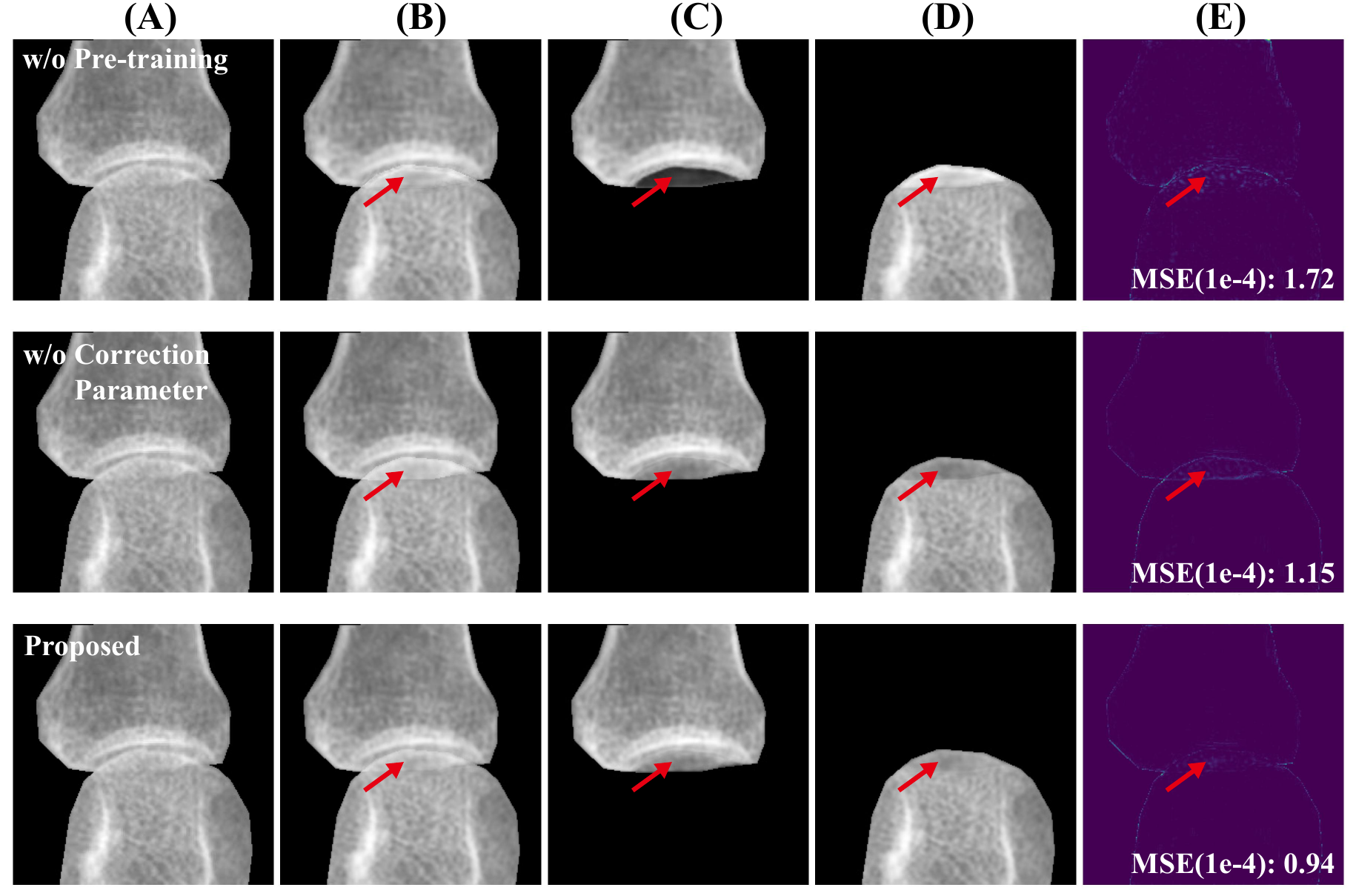}
  \centering
  \caption{Visualization results of ablation study. (A) Real Joint image; (B) Reconstructed Joint Image; (C) Upper Bone Layer; (D) Upper Bone Layer; (E) MSE Spectrum (A v.s. B).}
\label{fig: ablation_image}
\end{figure}

\subsection{Ablation Study}
We conducted ablation study centered on the integration of the correction parameter within the reconstructor, and synthetic image pre-training. Specifically, we examined three distinct framework configurations: the reconstruction function without correction parameter; training pipeline without the synthetic image pre-training; proposed framework.

As illustrated in Fig.~\ref{fig: ablation_plot}, Fig.~\ref{fig: ablation_image}, and Table \ref{tab: ablation}, The introduction of pre-training using synthetic images substantially enhances the quality of generated results. By establishing an initial model, the issues of erroneous generation due to network overfitting are markedly mitigated. Furthermore, the MSE distribution across overlap sizes has been significantly optimized, indicating that the framework performed more stably in different overlap sizes.
Additionally, the implementation of correction parameters leads to a considerable improvement in the reconstruction. In comparison to direct reconstruction methods, it effectively reduces brightness amplification in overlapping areas and improves texture synthesis. This enhancement is particularly notable in scenes with extensive overlapped regions, where the continuity and clarity of textures are significantly better. Moreover, the MSE distribution for occluded area sizes shows substantial improvement when compared to direct reconstruction techniques.
In conclusion, the integration of pre-training and correction parameters significantly enhances the stability and quality of the generated outputs, further underscoring the necessity of their introduction.

\begin{figure*}[!t]
  \includegraphics[width=\textwidth]{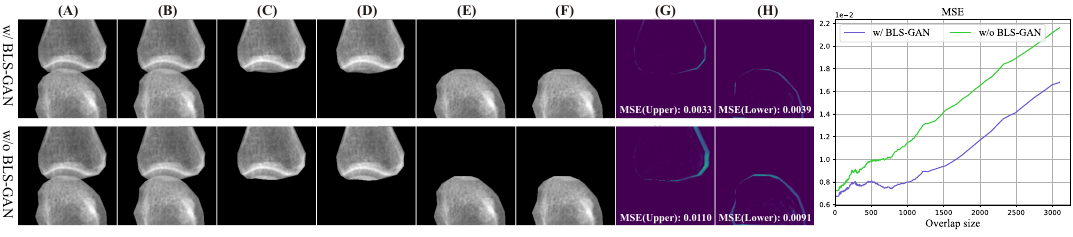}
  \centering
  \caption{Comparison of JSN progression quantification performance with and without bone layer separation. The \textbf{left} panel illustrates the visualization results, while the \textbf{right} panel depicts the curve of MSE across the overlap size. (A) Moving Joint Image; (B) Fixed Joint Image; (C) Moving Upper Bone; (D) Fixed Upper Bone; (E) Moving Lower Bone; (F) Fixed Lower Bone; (G) MSE Upper Spectrum (C v.s. D); (H) MSE Lower Spectrum (E v.s. F).}
\label{fig: JSN}
\end{figure*}

\subsection{Expert Assessments: Visual Turing Test}
We conducted a Visual Turing Test on three sets of 50 images each, comprising joint, upper bone, and lower bone images, with a real-to-fake ratio of 1:1, which informed to subjects. For the joint images, real images were paired with synthetic images used for pre-training. The upper and lower bone sets contained real and generated layer images. Four subjects with 12, 17, 26, and 30 years experiences as radiological technologist in the test, which lasted four hours.

The results in Table \ref{tab: turingtest} demonstrated that for the joint image sets, the scores of the four radiological technologists were considerable. But the combined metrics of accuracy, sensitivity, and specificity for the upper and lower image sets were around 0.5. This variability suggests that the synthesized joint images are not completely identical to the real ones, but the aggregated accuracy indicates their suitability as pre-training data. Additionally, the near-random ability of observers to distinguish real from generated images in the upper and lower sets suggests our method has effectively passed the visual Turing test.

\begin{table}[!t]
\caption{Visual Turing Test evaluation results over three images groups. Radiological technologist were tasked to label each set of images as real or fake.}
\centering
\resizebox{\linewidth}{!}{
    \begin{tabular}{llccccc}
        \toprule
        \multicolumn{2}{c}{} & R1 & R2 & R3 & R4 & Overall \\
        \midrule
        \multirow{3}{*}{\shortstack{Joint Image\\(Real \& Synthetic)}} & sensitivity & 0.76 & 0.36 & 0.24 & 0.40 & 0.44 \\
        & specificity & 0.52 & 0.40 & 0.20 & 0.44 & 0.39 \\
        & accuracy & 0.64 & 0.38 & 0.22 & 0.42 & 0.41 \\
        \midrule
        \multirow{3}{*}{\shortstack{Upper bone image\\(Real \& Generated)}} & sensitivity & 0.56 & 0.48 & 0.52 & 0.60 & 0.54 \\
        & specificity & 0.44 & 0.64 & 0.48 & 0.32 & 0.47 \\
        & accuracy & 0.50 & 0.56 & 0.50 & 0.46 & 0.51 \\
        \midrule
        \multirow{3}{*}{\shortstack{Lower bone image\\(Real \& Generated)}} & sensitivity & 0.64 & 0.60 & 0.44 & 0.36 & 0.51 \\
        & specificity & 0.60 & 0.44 & 0.64 & 0.36 & 0.51 \\
        & accuracy & 0.62 & 0.52 & 0.54 & 0.36 & 0.51 \\
        \bottomrule
    \end{tabular}
}
\label{tab: turingtest}
\end{table}

\subsection{Clinical Validation: JSN Quantification}
We conducted experiments to clinical performance on downstream tasks, JSN progression quantification, comparing the MSE differences between results with and without the introduction of our bone layer separation framework.

JSN is a crucial indicator for MSK diagnosis, especially for RA progression. \cite{wang2023deep} demonstrated that JSN can be quantified using deep registration to analyze changes between fixed (baseline) and moving (follow-up) images of finger joints. This method utilizes images and joint masks as inputs to produce registration parameters for JSN calculation. The MSE between registered and fixed images validates the registration results.

The MSE results show that the pipeline incorporating BLS-GAN achieves an MSE of 0.0088 ± 0.0118, which is notably lower than the MSE of 0.0103 ± 0.0133 observed in the pipeline without BLS-GAN 
(P value $<$ 0.0001, 95\% confidence interval: 0.001197 to 0.001828, Paired T-test).
As shown in Fig. \ref{fig: JSN}, the experimental outcomes further indicate that the introduction of the bone layer separation framework substantially improves both the accuracy and stability of deep registration in JSN quantification, particularly when overlap sizes are less than 1000 pixels. Although the MSE of our framework increases linearly with larger overlap sizes, it consistently remains lower than that of the deep registration method alone, thereby demonstrating the efficacy of ours in managing varying degrees of image overlap.

\section{Conclusion and Limitation}
This work initiated a challenging amodal completion scenario for medical images called bone layer separation, which aimed to address the impact of MSK joint bone overlap in conventional radiography. 
We implemented a GAN-based framework named BLS-GAN, which can provide a high-quality image with reasonable bone characteristics and texture. This framework is expected to eliminate the bone overlap in complex joints such as the wrist, hip, and knee, extending the application of automated quantitative methods to a broader range in conventional radiography.

This framework uses a unique reconstructor based on absorption-based imaging principles, reducing recurrent calculations in soft tissue and achieve high-quality reconstruction. The segmentation-based multi-channel supervisor network accurately distinguishes between overlapped and non-overlapped regions and verifies the authenticity of the generated images. Additionally, synthetic images pre-training enhances the stability of training process and generation.

The expert assessments and clinical validation demonstrated that the framework is capable of generating bone layer images with high clarity, exceptional stability, and a remarkable resemblance to real images. Additionally, the framework significantly enhances the accuracy and stability of downstream JSN quantification tasks. The introduction of our framework addresses the challenges of misalignment and instability caused by overlap in deep registration methods, thereby promoting broader adoption of JSN quantification using deep registration. Additionally, this advancement establishes a practical foundation for extending the method to more complex joints with intricate overlap and provides a solid technical basis for comprehensive, high-precision JSN quantification analysis. To the best of our knowledge, this study is the first application and exploration of amodal completion in conventional radiographs, enabling new developments in amodal completion in medical imaging.

Our current framework is designed to extract bone structures from raw joint radiographs, neglecting essential soft tissue information. While a correction parameter reduces recurrent soft tissue calculations in overlapped regions by adjusting brightness, it does not completely address soft tissue texture interference, which compromise image quality in both overlapped and non-overlapped regions. Future work will focus on accurately generating and differentiating soft tissue regions from bone layers, which is challenging due to the lack of positive samples (without bones) for supervision.

\bibliography{aaai25}

\end{document}